\documentclass[12pt,preprint]{aastex}

\newcommand\sfr{{\rm SFR}}
\newcommand\loh{{12+\log{\rm(O/H)}}}

\begin{document}

\shortauthors{Stanek et al.}

\shorttitle{No GRBs in Metal-Rich Galaxies}

\title{Protecting Life in the Milky Way: Metals Keep the GRBs Away}

\author{K.~Z.~Stanek\altaffilmark{1},
O.~Y.~Gnedin\altaffilmark{1},
J.~F.~Beacom\altaffilmark{2,1},
A.~P.~Gould\altaffilmark{1},
J.~A.~Johnson\altaffilmark{1},
J.~A.~Kollmeier\altaffilmark{1},
M.~Modjaz\altaffilmark{3},
M.~H.~Pinsonneault\altaffilmark{1},
R.~Pogge\altaffilmark{1},
D.~H.~Weinberg\altaffilmark{1}
}

\altaffiltext{1}{\small Dept.~of Astronomy, The Ohio State University, Columbus, OH 43210}
\altaffiltext{2}{\small Dept.~of Physics, The Ohio State University, Columbus, OH 43210}
\altaffiltext{3}{\small Harvard-Smithsonian Center for Astrophysics, Cambridge, MA 02138}

\email{kstanek@astronomy.ohio-state.edu}

\begin{abstract}

The host galaxies of the five local, $z\leq 0.25$, long-duration
gamma-ray bursts (GRBs 980425, 020903, 030329, 031203 and 060218),
each of which had a well-documented associated supernova, are all
faint and metal-poor compared to the population of local star-forming
galaxies.  We quantify this statement by using a previous analysis of
star-forming galaxies ($0.005 < z < 0.2$) from the Sloan Digital Sky
Survey to estimate the fraction of local star formation as a function
of host galaxy oxygen abundance.  We find that only a small fraction
($<25$\%) of current star formation occurs in galaxies with oxygen
abundance $12+\log{\rm(O/H)}<8.6$, i.e., about half that of the Milky
Way.  However, all five low-$z$ GRB hosts have oxygen abundance below
this limit, in three cases very significantly so.  If GRBs traced
local star formation independent of metallicity, the probability of
obtaining such low abundances for all five hosts would be $p\approx
0.1$\%.  We conclude that GRBs trace only low-metallicity star
formation, and that the Milky Way has been too metal rich to host long
GRBs for at least the last several billion years.  This result has
implications for the potential role of GRBs in mass extinctions, for
searches for recent burst remnants in the Milky Way and other large
galaxies, for non-detections of late radio emission from local
core-collapse supernovae, and for the production of cosmic rays in the
local Universe. Our results agree with theoretical models that tie
GRBs to rapidly spinning progenitors, which require minimal angular
momentum loss in stellar winds.  We also find that the isotropic
energy release of these five GRBs, $E_{\rm iso}$, steeply decreases
with increasing host oxygen abundance. This might further indicate
that (low) metallicity plays a fundamental physical role in the GRB
phenomenon, and suggesting an upper metallicity limit for
``cosmological'' GRBs at $\sim 0.15\;{\rm Z_{\odot}}$.

\end{abstract}


\section{Introduction}

Special circumstances are required to produce a long gamma-ray burst
(GRB).  While it has now been firmly established that these events
result from the death of very massive stars (e.g., Galama et al.~1998;
Stanek et al.~2003), there are two crucial features that distinguish
progenitors of long GRBs from the vast majority of other core collapse
supernovae.  First, there is strong evidence that GRBs are highly
beamed (e.g., Stanek et al.~1999; Rhoads 1999); second, the optically
detected supernovae are all Type Ic, lacking both hydrogen and helium
in their spectra (e.g., Stanek et al.~2003; Modjaz et al.~2006;
Mazzali et al.~2006; Mirabal et al.~2006).  This combination of
properties explains why they are so rare.  The presence of a jet
naturally implies rapid core rotation, which has been suggested by
theoretical studies (e.g., Woosley 1993); it is also easier for a jet
to penetrate the thin envelope of a star that has experienced strong
mass loss.  However, the extensive mass loss (increasing with
metallicity) required to produce Type Ic supernovae would normally
also cause extensive angular momentum loss. In this paper, we directly
assess whether such special circumstances exist by directly comparing
GRB hosts' metallicity to the metallicity of star forming galaxies in
the local Universe.

Studies of GRB hosts at $z\sim 1$ reveal that they are underluminous
compared to the general population of star-forming galaxies (e.g., Le
Floc'h et al.~2003; Fruchter et al.~2006), suggesting that GRBs occur
preferentially at low metallicities.  In our analysis we study the
five low redshift ($z\leq0.25$) GRBs, a complete sample of ``local''
bursts identified so far. In all cases these GRBs were followed by
well-documented supernovae.  This sample now includes GRB\,060218,
whose host is fainter than the Small Magellanic Cloud (Modjaz et
al.~2006).  There are several reasons why this sample is worth a
separate study.  Good abundance information exists for the hosts of
all five events, and it can be compared directly and using the same
techniques to the sample of local star-forming galaxies from the Sloan
Digital Sky Survey (SDSS) spanning approximately the same redshift
range. The highest redshift in the sample, $z=0.25$, corresponds to
look back time of $\sim 2/3$ of the age of the Earth, about the time
when life on Earth could be affected by GRB radiation. At these small
distances we might also see other impacts of GRBs, such as production
of cosmic rays and shell remnants.  With five well-studied events at
hand, for the first time there are enough data in this interesting
redshift range to make a direct and statistically significant
empirical study. This investigation complements the high-$z$ studies
and it {\it directly} addresses the properties of nearby GRBs and
their hosts, in case they are different.

The main result of our analysis is to show that the oxygen abundances
of the five hosts, which range from $\sim 0.1$ to $\sim 0.5$ of the
Solar value, are much lower than would be expected if local GRBs
traced local star formation independently of metallicity. We conclude
that GRBs are restricted to metal-poor stellar populations, in
agreement with recent theoretical models of their progenitors (e.g.,
Yon \& Langer 2005; Woosley \& Heger 2006), and that the Milky Way
and other large spirals have been too metal-rich to host GRBs for the
last several billion years (see also Langer \& Norman 2006). We
discuss several implications of this result.  We also find that the
$\gamma-$ray isotropic energy release, $E_{\rm iso}$, for these five
GRBs declines with increasing oxygen abundance of the host galaxy, and
suggest that the oxygen abundance threshold for a ``cosmological'' GRB
(visible at high redshifts) may be as low as 0.15 of the Solar value.

\section{Comparison of GRB Hosts with Local Star-Forming Galaxies}

Are the properties of long duration GRB hosts unusual compared with
the properties of normal galaxies in the local Universe?  We can
address this question by comparing the physical characteristics of
local GRB hosts directly to the same quantities for local galaxies in
the SDSS.

\begin{figure}[p]
\plotone{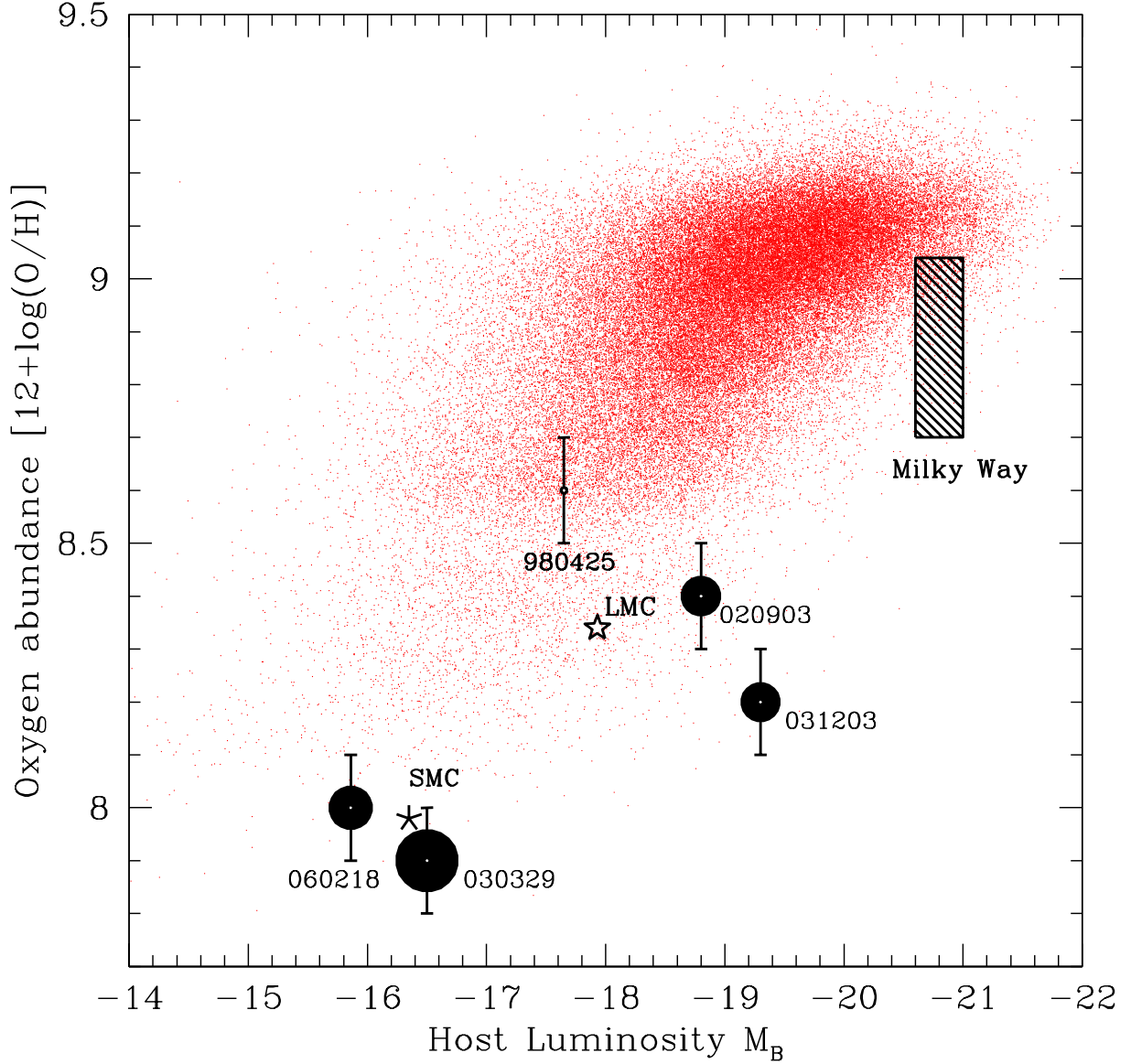}
\caption{Five low-$z$ GRB/SN hosts (filled circles) and local star
forming galaxies (small points: Tremonti et al.~2004; Tremonti 2006,
private communication) in the host luminosity-oxygen abundance
diagram. For comparison we also show the Milky Way, the LMC and the
SMC. It is clear that local GRB hosts strongly prefer metal-poor and
therefore low-luminosity galaxies. The circle areas for the GRB hosts
are proportional to the log of the isotropic $\gamma$-ray energy
release, $\log{E_{\rm iso}}$, for each burst, ranging from $\sim
1.0\times 10^{48}\;$erg for GRB 980425 to $\sim 2.0\times
10^{52}\;$erg for GRB 030329.  }
\label{fig_tremonti}
\end{figure}

Tremonti et al.~(2004) determine metallicities for a large sample of
SDSS galaxies from their spectra.  The redshifts of that sample are
restricted to $0.005 < z < 0.2$, providing a good comparison sample to
the local GRB hosts.  The metallicities are derived by a likelihood
analysis which compares multiple nebular emission lines ([\ion{O}{2}],
H$\beta$, [\ion{O}{3}], H$\alpha$, [\ion{N}{2}], [\ion{S}{2}]) to the
predictions of the hybrid stellar-population plus photoionization
models of Charlot \& Longhetti (2001).  A particular combination of
nebular emission line ratios arises from a model galaxy that is
characterized by a galaxy-averaged metallicity, ionization parameter,
dust-to-metal ratio, and 5500\AA\ dust attenuation.  For each galaxy,
a likelihood distribution for metallicity is constructed by comparison
to a large library of model galaxies.  The median of this distribution
is taken to be the galaxy metallicity, and the width of the
distribution is taken to be the error on the metallicity.
Figure~\ref{fig_tremonti} shows the galaxies from the extended sample
of ~73,000 star-forming SDSS galaxies studied by Tremonti et
al.~(2004) in the metallicity-luminosity plane.  We now add to this
diagram the local GRB hosts.

The large filled dots in Figure~\ref{fig_tremonti} mark the locations
of three previous GRB/SN hosts (SN\,1998bw, SN\,2003dh, SN\,2003lw)
with values of $M_B$ and $12 + \log{\rm (O/H)}$ taken mostly from
Sollerman et al.~(2005) (see Table~1 for references).  In addition, we
show the host of a very recent GRB\,060218/SN\,2006aj, whose host
galaxy has $12 + \log{\rm (O/H)}=8.0$ and sub-SMC luminosity (Modjaz
et al.~2006).  We also add a host of GRB\,020903 (Soderberg et
al. 2005; Bersier et al. 2006), which had a clear supernova signature
in its light curve, and was at fairly low redshift $z=0.25$. Oxygen
abundance for the host of GRB\,020903 has been recently measured by
Hammer et al. (2006).  The symbol areas for the GRB points in Figure~1
are scaled with isotropic $\gamma$-ray energy release $\log{E_{\rm
iso}}$ for each burst (see Table~1), ranging from $\sim 1.0\times
10^{48}\;$ergs for GRB 980425 to $\sim 2.0\times 10^{52}\;$erg for
GRB\,030329. There seems to be a progression of $E_{\rm iso}$ towards
lower energies with increasing oxygen abundance, which we will discuss
later in the paper.  As discussed in Sollerman et al., the applied
$R_{23}$ metallicity diagnostic (following Kewley \& Dopita 2002),
which employs emission line ratios of [\ion{O}{2}], [\ion{O}{3}] and
H$\beta$, is double-valued.  The degeneracy between the lower and
upper oxygen abundance branch can be broken by taking into account
other emission lines, e.g., [\ion{N}{2}].  For the host of
GRB\,030329, Sollerman et al.~(2005) could not break the degeneracy
due to the non-detection of [\ion{N}{2}], so they stated two possible
values for $12 + \log{\rm (O/H)}$, namely 8.6 and 7.9. Using the
published line ratios by Sollerman et al.~and Gorosabel et al.~(2005),
we consult Nagao, Maiolino \& Marconi (2006) who point to another
emission line diagnostic, namely [\ion{O}{3}]$\lambda
5007$$/$[\ion{O}{2}]$\lambda 3729$, that can give leverage in
distinguishing between the two branches. According to Nagao et al.,
when that ratio is above 2, the lower branch is favored, and we find a
value of 2.11 for that ratio. The lower value of $12 + \log{\rm
(O/H)}$ for the host of GRB\,030329 is also preferred by Gorosabel et
al. (2005) and seems more likely given its low luminosity---the upper
branch would predict a much brighter host galaxy according to the
luminosity-metallicity relationship. For GRB\,020903 Hammer et
al. (2006) derive $12 + \log{\rm (O/H)}=8.0$, using the effective
temperature method. That method has a significant offset from the
Kewley \& Dopita (2002) scale, so using the published values of line
fluxes in Table~1 of Hammer et al. we apply the prescription of Kewley
\& Dopita and obtain $12 + \log{\rm (O/H)}=8.4$. If we were instead to
use the formula from the very recent work of Kewley \& Ellison (2006)
to convert from the effective temperature method to the Kewley \&
Dopita method, we would add an offset of $+0.4\;$dex, in excellent
agreement with the previous value. We therefor adopt the final value
of oxygen abundance of $8.4$ for the host of GRB\,020903.

\begin{table}[h]
\caption{Properties of the Local GRBs/SNe and their Hosts}
\label{hosts}
\begin{tabular}{lccccc}
\hline
\hline
GRB   &   980425   & 020903  & 030329   &   031203  & 060218 \\
SN    &   1998bw   & \nodata & 2003dh   &   2003lw  & 2006aj \\
\hline
$z$ (redshift) &   0.0085$^{i}$  &  0.251$^{b,h}$  &    0.1685$^{i}$   &   0.1055$^{i}$  & 0.0335$^{f}$ \\
$E_{\rm iso}\; (10^{50}\;$erg) &  $0.010 \pm 0.002$$^{a}$ &  $0.28 \pm 0.07$$^{a}$  &  $180\pm 21$$^{a}$  &  $0.26\pm 0.11$$^{g}$  & $0.62\pm 0.1$$^{c}$ \\
$M_B$ (host)   &   $-$17.65$^{i}$ &   $-$18.8$^{b}$  &  $-$16.5$^{d}$  &   $-$19.3$^{g}$ & $-$15.86$^{f}$ \\
12+log[O/H] & 8.6$^{i}$ & 8.4$^{e,j}$  & 7.9$^{d,j}$  & 8.2$^{i}$  & 8.0$^{f}$
\end{tabular} \\
\tablenotetext{}{References: (a) Amati (2006); (b) Bersier et al.~(2006); (c) Campana et al.~(2006); 
(d) Gorosabel et al.~(2005); (e) Hammer et al.~(2006); (f) Modjaz et al.~(2006);
(g) Prochaska et al.~(2004); (h) Soderberg et al.~(2005); (i) Sollerman et al.~(2005); (j) this work }
\end{table}

For comparison, we also mark the locations of the Milky Way (including
a box to indicate the range due to the metallicity gradient, Carigi et
al. 2005; Esteban et al.~2005) and the Small and Large Magellanic
Clouds (Skillman, Kennicutt \& Hodge 1989) based on measurements of
individual HII regions (we use the values of $M_B$ from Arachnids
2005).  According to Esteban et al.~(2005), the value of $12 +
\log{\rm (O/H)}$ for the Solar circle is $8.70\pm 0.05$.  While in our
main analysis we directly compare nebular oxygen abundance between the
Tremonti et al.~sample and the GRB hosts, when referring to ``Solar
metallicity'', we adopt the Solar oxygen abundance of $12 + \log{\rm
(O/H)}=8.86$ (Delahaye \& Pinsonneault 2006).

It is indeed striking, that all of the local GRB hosts lie at
substantially lower metallicity than the vast majority of local
galaxies in the SDSS sample.  We quantify this result in the next
Section.

Note that we use the oxygen abundance values as derived from the
$R_{23}$ relationship by Kewley \& Dopita (2002), to be consistent
with the literature and to obtain the best relative values of the
oxygen abundance. Since different calibrations of the $R_{23}$
diagnostic have systematic differences of up to $0.2\;$dex at these
low abundances (see e.g., Nagao et al.~2006; Kewley \& Ellison 2006),
we decided to consistently use the same technique in comparing the GRB
hosts amongst themselves. In addition, the recent work by Kewley \&
Ellison (2006) shows that applying the method of Kewley \& Dopita
(2002) to the Tremonti et al. SDSS sample results in very good
agreement between the two methods, i.e. basically the Tremonti et al.
sample is effectively on the Kewley \& Dopita abundance scale.  We
should stress that our overall conclusion that the local GRBs only
occur in metal-poor galaxies does not depend on the exact choice of
$R_{23}$ calibration, because the GRB hosts so clearly happen only in
low-metallicity galaxies.

\section{Star Formation and Stellar Mass of GRB Hosts}

\begin{figure}[p]
\plotone{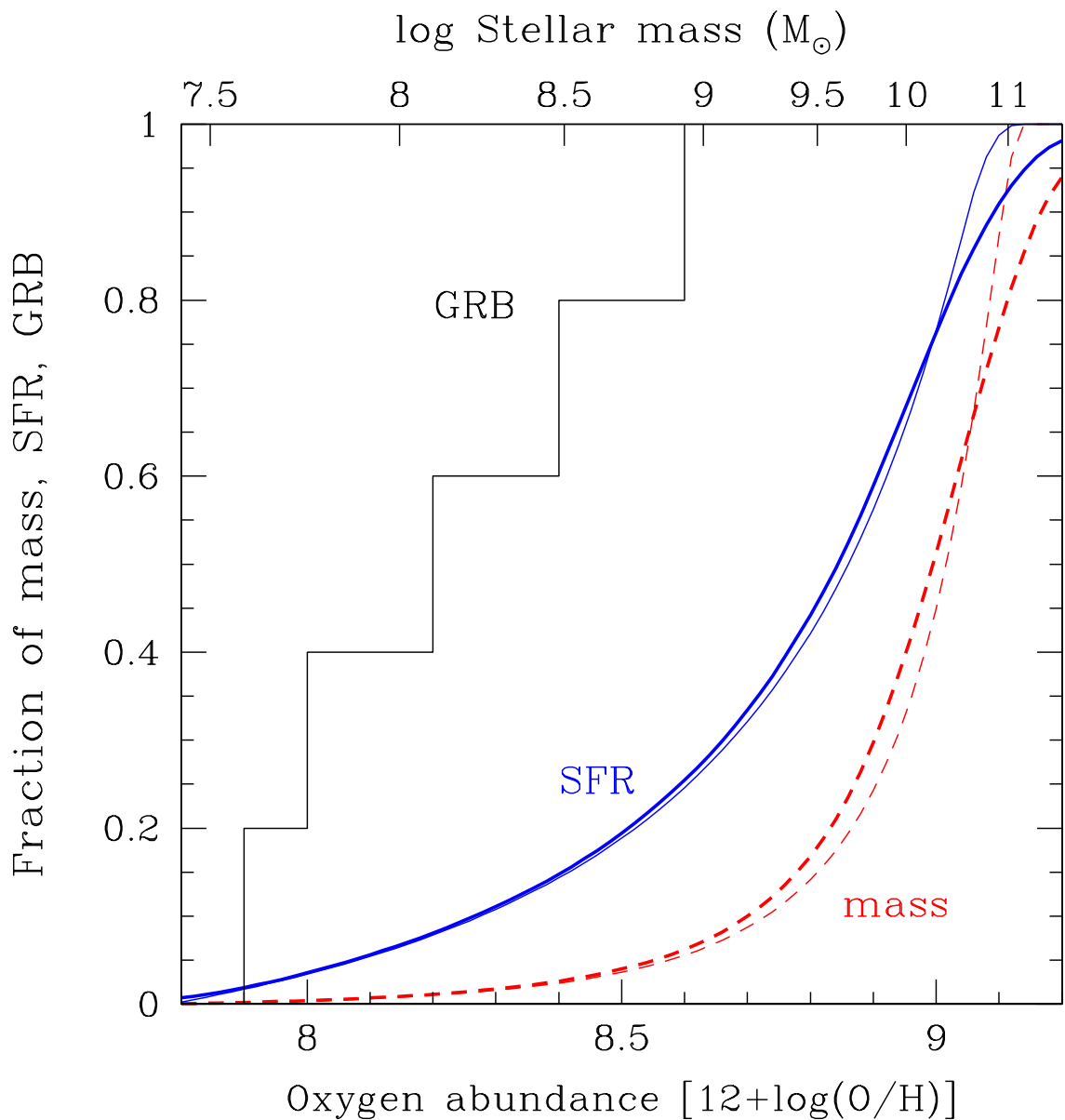}
\caption{Cumulative fractions of total star formation (solid lines)
  and total stellar mass (dashed lines) in late-type galaxies with the
  oxygen abundance below a given $12 + \log{\rm(O/H)}$.  Thick lines
  show the results of Monte Carlo realizations that include the
  estimated intrinsic scatter of the mass-metallicity and mass-SFR
  relations.  Thin lines show the results if there was no scatter.
  Solid histogram is the cumulative metallicity distribution of the
  five GRBs.  Top horizontal axis shows the corresponding scale of the
  galaxy stellar masses (Eq. 1).}
\label{fig_oleg}
\end{figure}

How improbable are the low oxygen abundances of the five low-redshift
GRB hosts? We test that under two ``null hypothesis'', one that GRBs
trace star formation, second that stellar GRBs trace star mass, in
both cases independently of metallicity.  We address this question
with a Monte Carlo test, by combining the Bell et al.  (2003)
measurement of the galaxy stellar mass function from the 2MASS and
SDSS surveys with the correlations of stellar mass with metallicity
and star formation rate ($\sfr$) measured for SDSS galaxies by
Tremonti et al. (2004) and Kauffmann et al. (2004), Brinchmann et
al. (2004), respectively.

The distribution of stellar masses, $M$, of galaxies in the local
Universe can be fit by a Schechter (1976) function, $\phi(M) dM
\propto \left({M/M^*}\right)^\alpha \exp{(-M/M^*)} dM$.  This
distribution is measured for galaxy masses $M > 10^9 \, M_\odot$.  We
have converted Bell et al.'s $M^*$ value from their ``diet Salpeter''
IMF to the Kroupa (2001) IMF used in the SDSS analysis, and we have
adopted the value of the Hubble constant $H_0 = 70$ km s$^{-1}$
Mpc$^{-1}$.  All galaxies in the sample have the characteristic mass
$M^* \approx 10^{10.85} \, M_\odot$ and the slope $\alpha = -1.1$,
while late-type only galaxies have $M^* \approx 10^{10.65} \, M_\odot$
and $\alpha = -1.27$.  The latter galaxies are closer match to the
star-forming galaxies considered in the other studies that we use
below.  This sample of late-type galaxies is also appropriate for
testing the hypothesis that GRBs trace star formation.

The mean stellar mass-metallicity relation of Tremonti et al. (2004)
has the form
\begin{equation}
  12 + \log{\rm (O/H)} = -1.492 + 1.847 \log{M} - 0.08026 \, (\log{M})^2,
  \label{eqn:oh}
\end{equation}
with the quoted scatter about the mean of 0.1 dex.  According to
Tremonti et al., this fit is valid in the stellar mass range $8.5 <
\log{M/M_\odot} < 11.5$.  We fit Brinchmann et al.'s (2004) relation
between $\sfr$ and $M$ by the broken power-law form
\begin{equation}
  \log{\sfr(M)} = 0.7 + \beta \, (\log{M} - 10.5),
  \label{eqn:sfr}
\end{equation}
with slope $\beta = +0.6$ for $\log{M} < 10.5$, where $\sfr$ is in
units of $M_\odot$ yr$^{-1}$.  Equation~(\ref{eqn:sfr}) is an eyeball
fit to the data in Fig.~17 of Brinchmann et al. (2004) in the mass
range $7 < \log{M/M_\odot} < 11$, from which we also estimate a
$1\sigma$ scatter of 0.3 dex about the mean relation.  At higher
masses, $10.5 < \log{M} < 11.5$, Brinchmann et al. find approximately
constant $\sfr$ ($\beta \approx 0$), while Kauffmann et al.'s (2004)
Fig.~7 indicates a significant downturn ($\beta \approx -0.6$).  In
the following, we consider the high-mass slope $\beta = -0.6$ as
standard and the other ($\beta = 0$) as a variation, and treat the
difference in inferred results as a systematic uncertainty associated
with the mass-metallicity modeling.

We use the above relations to calculate a fraction of stellar mass and
star formation rate contained in galaxies with metallicities below
those of the GRB hosts.  We generate Monte Carlo realizations of
$10^6$ galaxies with stellar masses drawn from the Bell et al. (2003)
mass function.  We have extrapolated this mass function below its last
measured point, down to $10^{7.4} \, M_\odot$, which corresponds to
the average metallicity $\loh \approx 7.8$, in order to include all
GRBs in our sample.  However, this is a conservative assumption since
without this extrapolation the mass and SFR fractions at low
metallicity would be even smaller.  For each galaxy, we draw a
metallicity and an $\sfr$ from the relations~(\ref{eqn:oh})
and~(\ref{eqn:sfr}), assuming log-normal scatter of 0.1 dex and 0.3
dex, respectively.  Note that we assume uncorrelated scatter between
these two quantities at fixed $M$.  To the extent that the
observational inputs are correct, this sample should have the same
joint distribution of mass, star formation rate, and metallicity as
real galaxies in the low-$z$ Universe.

The thick solid curve in Figure~\ref{fig_oleg} shows the cumulative
relation between star formation rate and oxygen abundance in the Monte
Carlo sample, i.e., the fraction of star formation in late-type
galaxies with oxygen abundance below the value on the $x$-axis.  The
thick dashed curve shows the corresponding cumulative relation for
stellar mass instead of star formation.  The thin solid and dashed
curves show the star formation and stellar mass relations,
respectively, if we ignore scatter and use just the mean
relations~(\ref{eqn:oh}) and~(\ref{eqn:sfr}).  In this case, the
fractions can be written analytically as $f_{\sfr} = \int_0^{M_{\rm
O/H}} \sfr (M)\, \phi(M) dM / \int_0^\infty \sfr (M)\, \phi(M) dM$ and
$f_{\rm mass} = \int_0^{M_{\rm O/H}} M \phi(M) dM / \int_0^\infty M
\phi(M) dM$, where $M_{\rm O/H}$ is the average mass corresponding to
the metallicity $\loh$ via relation (\ref{eqn:oh}).  Our Monte Carlo
sample without the intrinsic scatter gives identical results to these
analytical expressions.

\begin{figure}[p]
\plotone{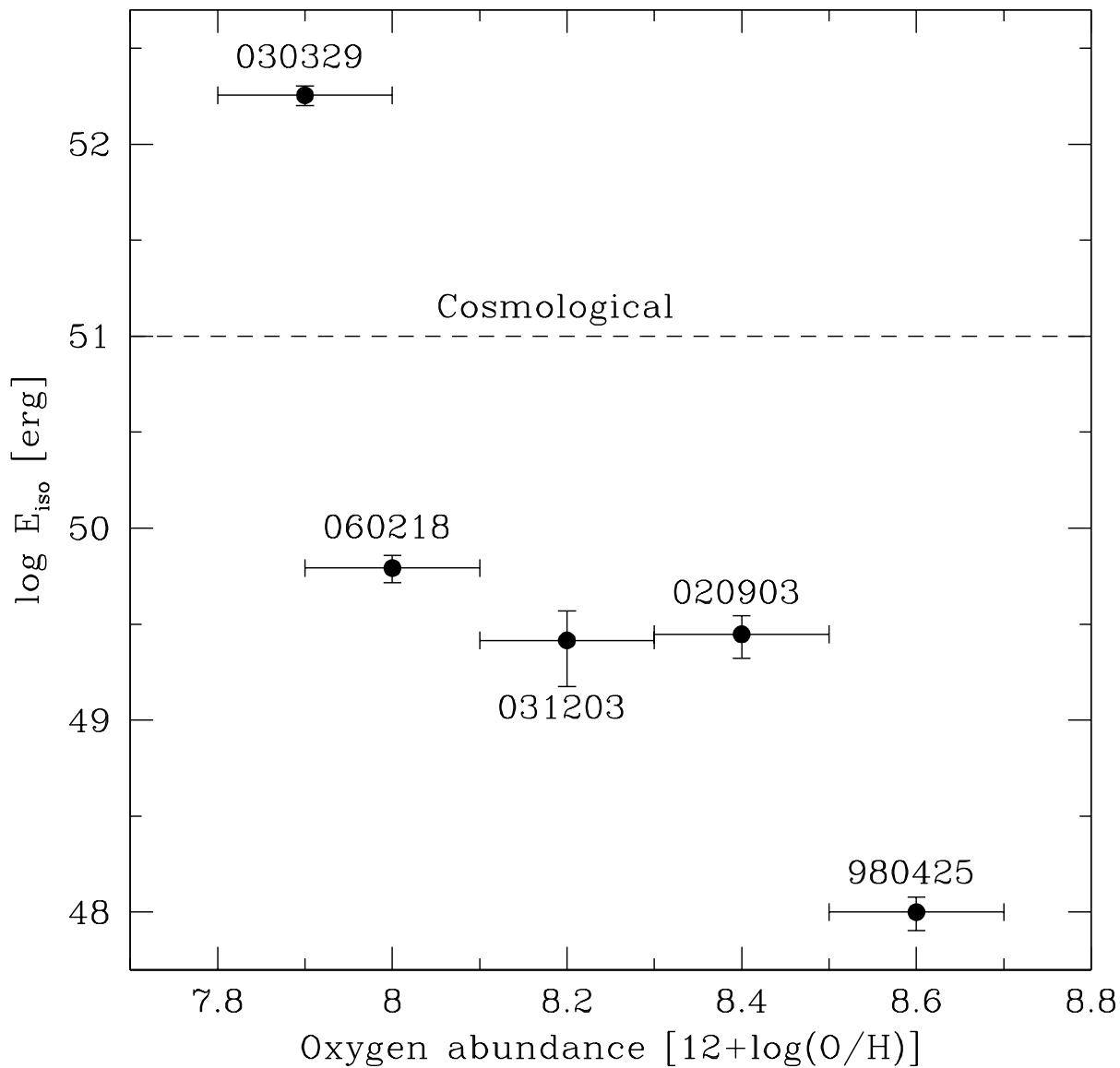}
\caption{Isotropic energy release in $\gamma$-rays, $E_{\rm iso}$, for
the five local GRBs plotted vs. the oxygen abundance of their hosts. A
strong dependence of $E_{\rm iso}$ on $12+\log{\rm (O/H)}$ seems to be
present, with a possible threshold for making ``cosmological'' GRBs at
$12+\log{\rm (O/H)}=8.0$, i.e., about $0.15$ of the Solar oxygen
abundance. With dashed line at $E_{\rm iso}=10^{51}\;$erg we indicate
the approximate limit for ``cosmological'' long GRBs (see Table~1 in
Amati 2006). }
\label{fig_gamma}
\end{figure}

The histogram in Figure~\ref{fig_oleg} shows the cumulative oxygen
abundance distribution of the five low-$z$ GRBs, which is clearly very
different from that of star-forming galaxies.  In order to quantify
the statistical significance of this discrepancy, we have generated
new $10^6$ trials of selecting five ``hosts'' randomly from the
metallicity distribution function given by the SFR fraction (thick
solid line).  To their chosen metallicities we add an estimated
observational error, assuming it to be log-normal with the standard
deviation of 0.1 dex.  The maximum abundance among the five selected
hosts satisfies $\loh \leq 8.6$ only $p_{\rm max} = 0.13\%$ of the
time.  We also find that the median abundance of the five hosts
satisfies $\loh \leq 8.2$ only $p_{\rm med} = 0.5\%$ of the time.
Note that the median test may be sensitive to our extrapolation of the
Tremonti et al. (2004) relation below the range $\loh \ga 8.5$
constrained by the data.  Had we not taken into account the scatter of
the mass-metallicity or mass-SFR relations, the resulting
probabilities would be even lower.  If we draw model galaxies from the
mass function of all (not only late-type galaxies), the probabilities
are at least a factor of 10 lower.  We have also used a standard
Kolmogorov-Smirnov test with a sample size $N=5$.  The KS probability
of the observed GRB metallicities being drawn from the SFR
distribution is 0.32\%, consistent with our Monte Carlo result, while
the probability of being drawn from the mass distribution is only
$0.008\%$.

Our results are not sensitive to the variation of the high-mass slope
of the mass-star formation rate relation.  If we take $\beta = 0$ at
$\log{M} > 10.5$, the probabilities change only slightly and shift
only towards smaller values.  The results of our models are summarized
in Table \ref{tab:monte}.

Finally, we consider the most conservative scenario that our oxygen
abundance determination of GRB hosts is systematically off by up to
0.2 dex with respect to Tremonti et al.'s values.  We add $+0.2$ dex
to the maximum and median GRB metallicities (now 8.8 and 8.4,
respectively) and recalculate the Monte Carlo probabilities.  These
new probabilities are of course not as small as for our fiducial
metallicities, but nevertheless low.  The maximum abundance is
satisfied only in 2\% of the cases and the median in less than 3\% of
the cases.  Note that we consider this arbitrary shift as an extreme
scenario and that we believe our GRB metallicities to be correct as
described in the previous Section and given in Table~1.

\begin{table}[t]
\caption{Monte Carlo Probabilities}
\label{tab:monte}
\begin{tabular}{lll}
\hline
\hline
 Model              & $p_{\rm max}$    & $p_{\rm med}$  \\
\hline
``standard''        &  0.0013   &  0.0051  \\
flat SFR            &  0.0012   &  0.0047  \\
$+0.2$ dex shift    &  0.020    &  0.028   
\end{tabular}\\ 
\tablenotetext{}{Probabilities of the GRB hosts tracing overall star
formation (independently of metallicity). See the paper for discussion.}
\end{table}

We conclude that even this fairly small sample of low-$z$ GRB hosts is
sufficient to show that GRBs do not trace the overall star formation
in the local Universe (and do no trace mass at extremely high
confidence).  Instead, GRBs arise preferentially in the lowest
metallicity systems.  In Figure~\ref{fig_tremonti}, it is striking
that GRB 031203, which has the brightest host galaxy, resides in a
system that is extremely metal-poor compared to other galaxies of its
luminosity.  Equally intriguing is the trend for brighter GRBs to
occupy the lowest metallicity hosts.  Figure~\ref{fig_gamma}
illustrates this point directly, plotting the isotropic $\gamma-$ray
energy release $E_{\rm iso}$ against $\loh$ (for an earlier, indirect
attempt to correlate $E_{\rm iso}$ with host metallicity see Fig.1 in
Ramirez-Ruiz et al. 2002).  The low energies of the low-$z$ GRBs have
been discussed by many authors ever since the discovery of GRB 980425.
In principle the low values of $E_{\rm iso}$ could arise from beaming
effects, with the proximity of the bursts allowing us to see them
further off-axis, but Cobb et al. (2006) argue persuasively against
this interpretation.  If $E_{\rm iso}$ is reasonably representative of
the true energetics of these low-$z$ GRBs, then Figure~\ref{fig_gamma}
suggests that there may be a threshold for producing truly
``cosmological'' GRBs that are bright enough to be seen to high
redshift, at an oxygen abundance $\loh \sim 8.0$, roughly 0.15 of the
Solar abundance. We caution that this trend is rather speculative
given the current data, unlike the main result of our of paper, i.e.,
that local GRBs occur only in metal-poor galaxies.

\section{Discussion}

Our findings for local GRBs are in qualitative agreement with the
studies showing that high-redshift GRBs reside in underluminous
galaxies (e.g., Le Floc'h et al.~2003; Fruchter et al.~2006).  The
advantage of studying the local sample is that we can focus directly
on metallicity, which appears to be the critical physical parameter,
and we can compare the GRB host metallicities to those measured in
local star-forming galaxies. The arguments in \S 2 and \S 3 indicate
that long GRBs occur only in low metallicity environments, and
therefore do not occur in ``normal'' galaxies that are comparable to
the Milky Way in mass and metallicity.  This has a number of
implications, some of which have been discussed independently by
Langer \& Norman (2006) based on an entirely different line of
argument involving higher-$z$ GRBs.

Our results agree well with recent theoretical work on GRB
progenitors. The collapsar model, where the GRB is created by an
accretion disk around a rotating black hole, requires the core angular
momentum of the progenitor to be dynamically important at the time of
collapse.  This requirement sets severe limits on core angular
momentum loss, which would normally accompany the substantial mass
loss associated with the Wolf-Rayet stars thought to be the
progenitors of typical Type Ic supernovae. Two viable channels have
been proposed, both of which avoid the red supergiant phase.  First,
interactions with a close binary companion can strip the envelope too
rapidly for the core to be spun down (see Podsiadlowski et al.~2004
for a detailed discussion).  Second, a single star that rotates
rapidly enough can experience fully mixed evolution (Yoon \& Langer
2005; Woosley \& Heger 2006) and avoid the red supergiant phase
entirely.  The latter mechanism also avoids core contraction during
the hydrogen and helium burning phases, which would further shield the
core from angular momentum loss associated with magnetic fields
(Spruit 2002; however, see Denissenkov \& Pinsonneault 2006). With
either of these mechanisms, however, GRBs would not be expected for
high iron abundances because of strong mass and angular momentum loss
during either the main sequence or the Wolf-Rayet phase (Heger \&
Woosley 2002).  Yoon \& Langer (2005) and Woosley \& Heger (2006)
estimate that an iron abundance of about 0.1 Solar is a maximum
threshold for such a mechanism.  The existence of a strong metallicity
threshold therefore provides support for recent theoretical models of
the formation of long GRBs, and with better statistics we may be able
to distinguish between the different formation channels.

The iron abundance is more important than the oxygen abundance in this
regard because iron provides much of the opacity for radiation-driven
stellar winds (e.g., Pauldrach, Puls, \& Kudritzki 1986). Our use of
oxygen as a proxy for metallicity may therefore underestimate the
significance of the abundance trends that we observe.  The earliest
generations of stars are known to be enhanced in [O/Fe] relative to
the Solar mixture (Lambert, Sneden \& Ries 1974). It is therefore
likely that the GRB host galaxies are even more iron-poor than they
`are oxygen-poor. The specific frequency of Wolf-Rayet stars relative
to O stars is an order of magnitude higher in high metallicity spirals
than it is in systems such as the SMC (Maeder \& Conti 1994). Since
normal Type Ic supernovae are associated with Wolf-Rayet progenitors,
the low metallicity of the five local GRB hosts is even more
significant, as Type Ic supernovae in general trace metal-rich star
formation.

An upper limit on metallicity for long GRBs has a number of other
consequences.  GRBs are unlikely to be a source of cosmic rays in the
Milky Way (a possibility discussed by, e.g., Dermer 2002), and they
can play only a limited role in cosmic ray production in the
low-redshift Universe. Searches for GRB remnants in nearby large
galaxies (e.g., Loeb \& Perna 1998) are expected to yield few, if any,
detections.  We also argue that asymmetric supernovae remnants
observed in the Milky Way did not result from recent GRB explosions
(e.g., Fesen et al.~2006; Laming et al.~2006).  It also follows that
late-time non-detections of radio emission from local core-collapse
supernovae (e.g., Soderberg et al.~2006), while providing interesting
constraints on their physics, do not provide information on the
beaming or circumstellar environments of GRBs.  These core-collapse
SNe are most likely located in higher metallicity galaxies that are
unlikely to produce a GRB.

A GRB occurring in the last billion years within a few kiloparsecs
from Earth has been invoked as a possible cause for a mass extinction
episode (e.g., Thomas et al.~2005a,b). Our results make this scenario
most unlikely---by the time the Earth formed, the Milky Way disk was
already too metal-rich to host a long GRB.  SN\,1998bw/GRB\,980425,
the only local event to happen in a fairly metal-enriched galaxy, was
also by far the weakest localized GRB ever, with at least 10,000 times
lower energy than a typical $z\sim 1$ GRB. As such, it would not cause
mass extinction at several kpc from Earth.  The same can be said about
short GRBs, which are not only less frequent than long GRBs (e.g.,
Kouveliotou et al.~1993), but also less energetic and less beamed
(e.g., Grupe et al.~2006; Panaitescu 2006). Short GRBs are also not
concentrated to star-forming regions, thus on average they are much
further away from any life-hosting planets (e.g., Bloom \& Prochaska
2006).  In addition, planet-hosting stars are on average even more
metal rich than the Sun (e.g., Santos, Israelian \& Mayor 2004),
making long GRBs an even less likely source of life extinction events
in the local Universe. So to finish with a bit of good news, we can
probably cross GRBs off the rather long list of things that could
cause humankind to ``join the dinosaurs'' on the extinct species list.

\acknowledgments 

We thank Christy Tremonti for making her extended dataset available to
us. We are grateful to Lisa Kewley for sharing with us the results of
her recent work in advance of publication. We thank the anonymous
referee, David Bersier, Andy Fruchter, Norbert Langer, Bohdan
Paczynski, Enrico Ramirez-Ruiz and Christy Tremonti for useful
comments on an earlier version of this manuscript.  We would also like
to thank the participants of the morning ``Astronomy Coffee'' at the
Department of Astronomy, The Ohio State University, for the daily and
lively astro-ph discussion, one of which prompted us to investigate
the problem described in this paper. JFB is supported by NSF CAREER
grant No. PHY-0547102. AG. and JAK were supported by grant AST-0452758
from the NSF.

\end{document}